\documentclass[twocolumn,english,aps,prb,twocolum,superscriptaddress,natbib,bibnotes,amsmath,amssymb,floatfix,groupedaddress,footinbib]{revtex4-2}
\pdfoutput=1
\usepackage[colorlinks=true,citecolor=blue,linkcolor=magenta]{hyperref}

\usepackage[markup=nocolor, authormarkupposition=left]{changes} 

\usepackage{soul}
\usepackage[utf8]{inputenc}
\usepackage[english]{babel}
\usepackage{amsmath,amsfonts,amssymb}
\usepackage[T1]{fontenc}
\usepackage{url}
\usepackage{amsmath}
\usepackage{siunitx}
\usepackage[version=4]{mhchem}
\usepackage{amsfonts}
\usepackage{amssymb}
\usepackage{epstopdf}
\usepackage{graphicx}
\graphicspath{{./Figures/}}

\usepackage{changes}
\usepackage{titlesec}

\newcommand{\red}[1]{\textcolor{black}{#1}}

\begin{document}

\title[Article Title]{Towards Terabit/$\lambda$/s Multidimensional Silicon Photonic Engine}

\author{Hao Chen$^{1,2,\ddagger}$, Zengqi Chen$^{1,2,\ddagger}$, Wu Zhou$^{1}$, Kaihang Lu$^{1}$, Mingyuan Zhang$^{1}$, Yuxiang Yin$^{1}$, Yiou Cui$^{1}$, Chaoran Huang$^{3}$, Pui-In Mak$^{4}$, $\&$ Yeyu Tong$^{1,2,*}$}

\affiliation{$^1$Microelectronic Thrust, The Hong Kong University of Science and Technology (Guangzhou), 511453, Guangzhou, Guangdong, PR China}
\affiliation{$^2$Guangdong-Macao Joint Laboratory for Modular Chip Design and Testing, The Hong Kong University of Science and Technology (Guangzhou), 511453, Guangzhou, Guangdong, PR China}
\affiliation{$^3$Department of Electronic Engineering, The Chinese University of Hong Kong, Hong Kong, PR China}
\affiliation{$^4$Faculty of Science and Technology, University of Macau, Macau, PR China}
\affiliation{{$^\ddagger$}These authors contributed equally to this work.} 
\affiliation{{$^*$}Corresponding authors: \href{mailto:yeyutong@hkust-gz.edu.cn}{yeyutong@hkust-gz.edu.cn}}

\maketitle

\noindent\textbf{\red{Increasing artificial intelligence (AI) workloads drive co-packaged optics (CPO), which integrates optical engines with electronic components. Optical interconnects can extend transmission distances and reduce latency, allowing distributed clusters in AI factories to operate as a unified computational unit. However, escalating data throughput necessitates greater parallelization of light within ultracompact form factors while maintaining stringent energy efficiency and latency constraints. Here, we present a multidimensional silicon photonic engine that achieves a communication capacity exceeding 1.8 terabit/$\lambda$/s. By monolithically integrating transceivers, spatial and polarization (de)multiplexers, and optical signal processors on a single chip, we eliminate bulky discrete (de)multiplexers and power-hungry digital signal processing (DSP). In experiments, the photonic engine can be self configured to identify two, four, or six concurrent spatial and polarization channels per fiber while mitigating dynamic channel crosstalk. Compared with the state-of-art DSP, our approach achieves >5,000-fold reductions in both power consumption and processing latency at a MIMO processing order of six. Furthermore, we demonstrate full‑duplex, modulation‑format‑transparent inter‑chip communication over 300-meter fiber. These results represent a paradigm shift for optical engines in future high‑performance computing and AI-driven data centers.}}

\section*{Introduction} 

\begin{figure*}
  \includegraphics[width=0.92\linewidth]{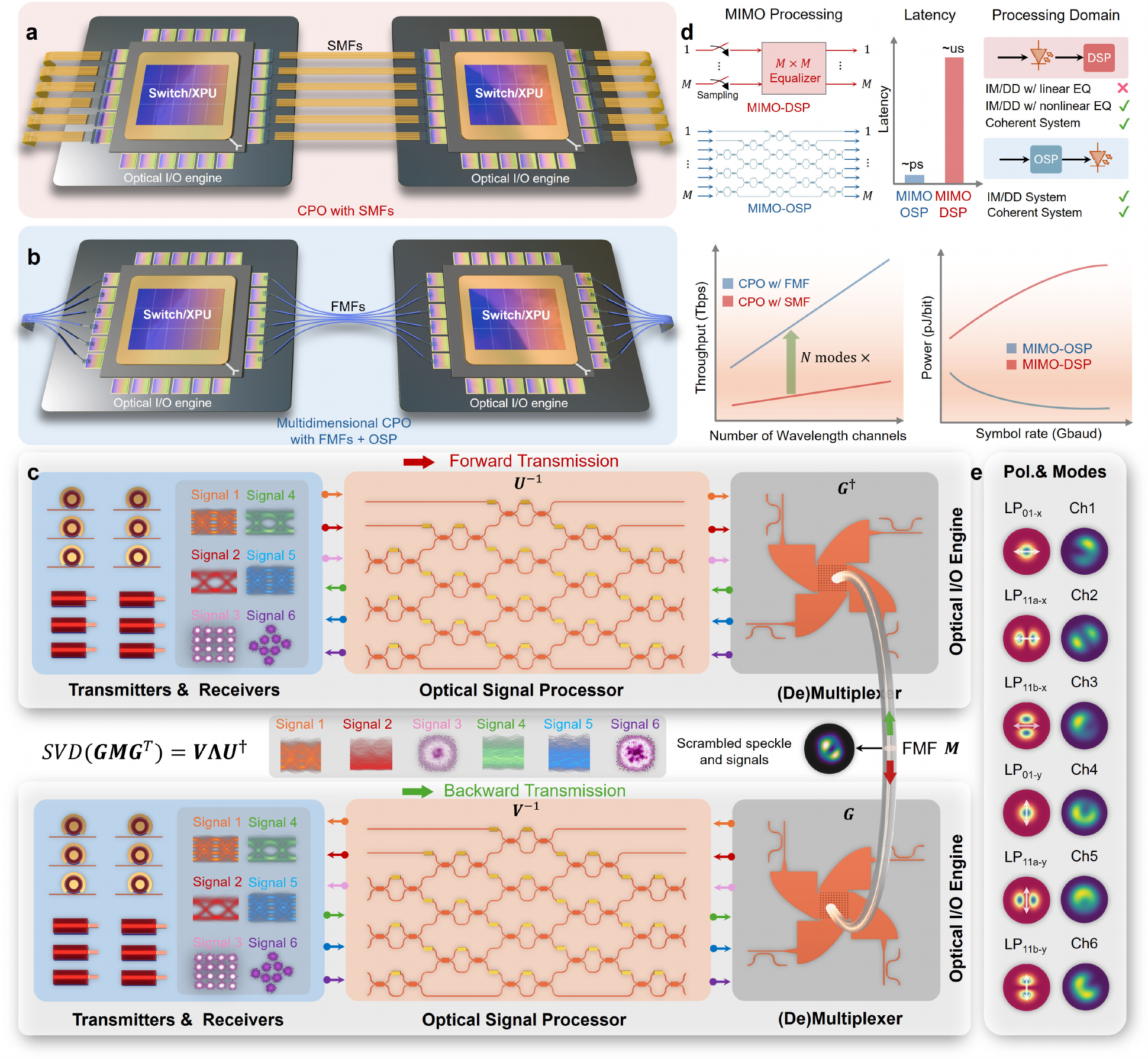}
  \caption{\red{\textbf{Few-mode fiber based optical engine.} \textbf{a $\&$ b} Artistic representation of CPO that integrates optical chiplets directly with electrical chips (Switch, XPUs) in the same package. "XPUs" refers to abstraction including various computational architectures, including CPUs, GPUs, FPGAs, and other accelerators. Single-mode fibers (SMFs) are employed currently to bridge the optical engines. In this work, the multidimensional optical link leverages various dimensions of light within a few-mode fiber (FMF), including mode and polarization. Given the enhanced communication capacity per fiber cross section, the number of optical cables can potentially be reduced. \textbf{c} Schematic illustration of the optical engine facilitating multidimensional and full-duplex inter-chip optical communications. The optical transceivers, multiplexers/demultiplexers (MUX/DEMUX), and optical signal processors are monolithically integrated into the same photonic chip. The optical fiber is directly interfaced with the optical engine using an out-of-plane coupling method. Silicon micro-ring modulators (MRMs) and germanium-on-silicon photodiodes (PDs) serve as transmitters and receivers. The MUX and DEMUX for FMF employ diffraction grating antennas that accommodate mode and polarization diversity. The dynamic beating of six orthogonal spatial and polarization modes in a two-mode fiber induces scrambled speckle patterns, mixed signal eye diagrams and constellation diagrams. To mitigate that, two tunable Mach-Zehnder interferometer (MZI)-based optical signal processors are employed to implement singular value decomposition (SVD) based mulitple in multiple out (MIMO) optical signal processing (OSP). High-speed signals can be recovered with transparency to modulation format and data rate. In this work, three data channels are allocated for forward transmission and the other three are utilized for backward transmission. \textbf{d} Figure of merits comparison between MIMO digital signal processing (DSP) and OSP equalizer (EQ), including modulation format compatibility, throughput against number of wavelength channels, and energy efficiency scaled with symbol rates. $N$ is the number of incorporated spatial and polarization channel. \textbf{e} Simulated intensity profiles of the multiplexed fiber mode set that are launched by the grating antenna, including basic linearly-polarized (LP) mode set, and randomly rotated channel set identified from self configuration. Due to the spatial and polarization diversity, the intensity profiles of these channels exhibit a speckle-like pattern; however, they remain orthogonal to one another. }}
  \label{fig:1}
\end{figure*}

\red{The rapid progress of generative artificial intelligence (AI) has created an insatiable demand for data throughput \cite{mahajan2021co, tan2023co}, placing unprecedented strain on interconnect fabrics in terms of bandwidth density, energy efficiency, and latency to effectively accommodate the computational needs and massive datasets associated with AI model training and inference. Optical interconnects present a promising alternative to copper-based connections. By utilizing light for data transmission, higher bandwidth, longer reach, and lower latency \cite{demir2014galaxy} can be provided for enabling large-scale distributed AI clusters to function as a cohesive computational unit. However, conventional pluggable optical modules are approaching fundamental limitations in these metrics, driving the development of co-packaged optics (CPO) \cite{mahajan2021co, wade2020teraphy, fathololoumi20201, minkenberg2021co}, where optical engines (modulators, detectors) and electronic components (switch, XPUs) are integrated within the same package, as illustrated in Fig. \ref{fig:1}a. Silicon photonics is the one of the favoured platforms for CPO due to its compatibility with CMOS fabrication and high integration density. Current designs rely on wavelength division multiplexing (WDM) to scale capacity, typically using 4 to 16 wavelength channels per fiber \cite{sun2020teraphy, mahajan2021co}. Yet the symbol rate per channel is approaching 200 Gbaud, which is close to their physical limit of silicon modulators \cite{li2023integrated, steckler2025monolithic, lu2025whispering}. Further scaling of wavelength channels is hindered by the cost and power consumption of dense WDM laser sources, as well as the stringent performance demands of integrated wavelength multiplexers (MUXs) and demultiplexers (DEMUXs).}


\red{To overcome these constraints, the transverse wavefront of light can be utilized to include extra polarization and spatial modes \cite{luo2014wdm, yang2022multi}. This goes beyond the currently employed single-polarized quasi-Gaussian mode in single-mode fibers (SMFs). Polarization division multiplexing (PDM) and mode division multiplexing (MDM) have been extensively studied in fiber communications \cite{puttnam2021space, richardson2013space}, such as multimode fibers (MMFs) \cite{sillard201650} or few-mode fibers (FMFs) \cite{kitayama2017few} , but their integration into compact, scalable optical engines remains elusive. Traditional spatial-mode multiplexers, such as multi-plate light conversion, photonic lanterns, and three-dimensional waveguide couplers are too bulky for chip-scale deployment \cite{rademacher2021peta, fontaine2012geometric, gross2014three, wan2024multidimensional}. Moreover, dynamic inter-channel crosstalk caused by fiber bending, twists, and manufacturing defects presents a fundamental challenge \cite{palmieri2025mode}. In long-haul systems, this is mitigated by multi-input multi-output (MIMO) digital signal processing (DSP) with coherent detection \cite{savory2010digital}, but the associated power consumption, latency, and hardware cost are prohibitive for short-reach interconnect applications, particularly in AI infrastructure where every milliwatt and nanosecond matters \cite{hout2024transmission, ryf2012mode, ip2015sdm}. Recent advances in programmable photonic integrated circuits (PICs) have demonstrated optical signal processing capable of significantly reducing electronic power and latency \cite{bogaerts2020programmable, shastri2021photonics, shen2017deep, huang2021silicon, shibahara2025spatial, choutagunta2019adapting}. For example, on-chip interferometric meshes have been used for singular value decomposition to demultiplex modes in free-space, mobile, and on-chip systems \cite{fontaine2012efficient, annoni2017unscrambling, tang2018reconfigurable, wu2023chip, milanizadeh2022separating, zhang2024system, jian2025programmable}. A pair of integrated photonic processors can be self configured to identify optimal spatial mode set without prior knowledge of the free space system \cite{miller2013establishing, seyedinnavadeh2024determining, miller2013self}. Our own prior work \cite{lu2024empowering} showed that a programmable PIC interfaced with a FMF can adaptively descramble fiber modal crosstalk. Nevertheless, a universal, compact, low-power, and low-latency multidimensional optical engine for inter-chip communication has yet to be realized.}

\red{In this study, we report the first demonstration of a multidimensional silicon photonic engine that achieves a communication capacity exceeding one terabit per second per wavelength. Monolithic integration of high‑speed transceivers, spatial and polarization (DE)MUXs, and a SVD‑based optical signal processor on a single chip enables up to six concurrent orthogonal physical channels to be automatically identified through self‑configuration across a few‑mode fiber linking two photonic engines. This approach eliminates the need for bulky discrete optical (DE)MUXs and power‑intensive DSP, both prohibitive for short-reach interconnects. At a MIMO order of six, the proposed scheme yields >5,000‑fold reductions in both energy consumption and processing latency compared with state-of-the-art MIMO DSP. Full‑duplex, modulation‑format‑ and data‑rate‑transparent inter‑chip communication over 300 meters is demonstrated experimentally. Integrating multiple dimensions of light into a compact, low-power optical engine has the potential to enhance fiber throughput, reduce connector congestion, and streamline system integration. These findings outline a promising pathway for high-throughput short-reach optical interconnects using few-mode fibers, advancing next-generation AI infrastructure and data centers.}

\section{Operation Principle} 

The proposed optical engine is schematically depicted in Fig. \ref{fig:1}b. Compared with SMF-based solutions, multidimensional multiplexing offers enhanced data rate capacity per fiber cross-section, thus can potentially reduce the cable counts substantially. The detailed photonic engine is illustrated by Fig. \ref{fig:1}c. All the optoelectronic components are integrated on chip, including transceivers, optical signal processors, and (De)MUXs. In our work, compact and wavelength-selective silicon microring modulators (MRMs) and germanium-on-silicon photodiodes (PDs) are employed as transmitters and receivers respectively. To launch or receive various spatial and polarization beams, polarization- and mode-diversity grating couplers (GCs) are employed as MUX/DEMUX for fibers \cite{wohlfeil2016two, tong2019efficient, zhou2024ultra} with the transmission matrices denoted by $\mathbf{G}^{\dagger}$ and $\mathbf{G}$, respectively. The MUX/DEMUX is based on out-of-plane chip-fiber coupling that can support the complete set of six orthogonal polarization and modes in a two-mode fiber \cite{kitayama2017few}. The orthogonal basis vectors in a two-mode fiber can be described by six linearly polarized (LP) modes including $\text{LP}_{01x/y}$, $\text{LP}_{11ax/y}$, and $\text{LP}_{11bx/y}$ as presented by Fig. \ref{fig:1}e. Detailed design principle and simulation results of the grating coupler are included in the Supplementary Note S1.

The scrambled speckle and mixed signal diagrams shown in Fig. \ref{fig:1}c are caused by the random fiber transmission matrix $\mathbf{M}$. To solve this issue, MZI-based reconfigurable optical mesh is employed between the transceivers and (DE)MUXs on each photonic chip. In spite of various distortions, including mode dependent loss and crosstalk, the grating-fiber-grating transmission matrix $\mathbf{GMG}^{\dagger}$ can always be factorized by three simpler matrices $\mathbf{V\Lambda U}$ through singular value decomposition. As each photonic engine contains an unitary optical mesh which can be trained respectively to $\mathbf{U}^{-1}$ and $\mathbf{V}^{-1}$, the full chip-fiber-chip system can thus be annotated by 

\begin{equation}
\begin{split}
\sum\limits_{i=1}^{6}|\phi_{Oi}\rangle
&= \mathbf{V}^{-1} \mathbf{GMG}^{\dagger} \mathbf{U}^{-1} \sum\limits_{i=1}^{6}|\phi_{Ii}\rangle \\
&=\mathbf{V}^{-1} \mathbf{V} \mathbf{\Lambda UU}^{-1} \sum\limits_{i=1}^{6}|\phi_{Ii}\rangle \\
&= \mathbf{\Lambda} \sum\limits_{i=1}^{6}|\phi_{Ii}\rangle,
\label{eq1}
\end{split}
\end{equation}

\noindent where $\sum\limits_{i=1}^{6}|\phi_{Oi}\rangle$ and $\sum\limits_{i=1}^{6}|\phi_{Ii}\rangle$ represent the six modulated complex optical waves on two optical engines shown in Fig. \ref{fig:1}c. Hence, after proper configuration of the optical signal processors to perform MIMO-OSP, six concurrent data channels between two photonic engines can be directly bridged using a single FMF. Transmission of the six spatial and polarization channels are determined by the singular values of diagonal matrix $\mathbf{\Lambda}$. In this demonstration, the OSP MZI meshes were utilized on two separate chips. In future work, the SVD optical mesh may also be implemented exclusively on one receiver side. It is worthwhile to mention that differential group delay compensation is not implemented here, due to the short inter-chip communication distance using graded-indexed FMF. \red{Since the integrated MUX/DEMUX has eight fundamental-mode waveguides, which can be seamlessly connected with a reconfigurable Reck interferometric mesh with a matrix dimension of eight \cite{reck1994experimental}. Hence, when the optical engine operates with a two-mode circular-core fiber that supports up to six orthogonal spatial and polarization modes, the on-chip network preserves two redundant ports for potential future use.}

The comparison of MIMO-OSP and MIMO-DSP are shown in Fig. \ref{fig:1}d. The primary distinction lies in the signal processing domain, specifically whether processing occurs before or after the photodiodes. The presence of nonlinear distortions resulting from square-law photodetection in intensity modulation and direct detection (IM/DD) systems limits the efficacy of linear equalization in compensating for these nonlinear impairments. More sophisticated nonlinear equalizers or neural networks are thus needed for IM/DD scheme \cite{zhu2021dd, rajbhandari2019neural, jian2025programmable}. Coherent detection with a full-field recovery can maintain the system linearity to reduce MIMO processing complexity \cite{arik2014mimo}. However, the hardware costs and DSP requirements for coherent communication complicate the implementation. On the contrary, MIMO-OSP is modulation format and data rate transparent \cite{sacchi2025integrated}, meaning it can seamlessly work for the prevalent and low-cost IM/DD scheme utilized in current short-reach communication systems. The differing signal eye diagrams and constellation diagrams presented in Fig. \ref{fig:1}c illustrate that MIMO-OSP hardware does not require modifications to communication protocols, including symbol rates and modulation formats. Moreover, MIMO‑DSP introduces processing latencies exceeding microseconds, even without coherent overhead. MIMO‑OSP, by contrast, achieves processing at the speed of light through a simple forward transmission through the reconfigurable mesh, reducing latency to picoseconds. Compared with SMF-based optical engines, FMF-based design with MIMO-OSP enhances data throughput by incorporating additional spatial and polarization modes, as shown by Fig. \ref{fig:1}d. In addition, energy consumption per bit decreases as the symbol rate increases for MIMO-OSP. This is because each symbol requires less energy from the total power consumed by the reconfigurable optical processors. As optical engine continues to evolve, multidimensional multiplexing and MIMO-OSP emerge as a promising approach to introduce greater parallelism of light without issues raised in energy efficiency and interconnect latency.

\begin{figure*}
  \includegraphics[width=1\linewidth]{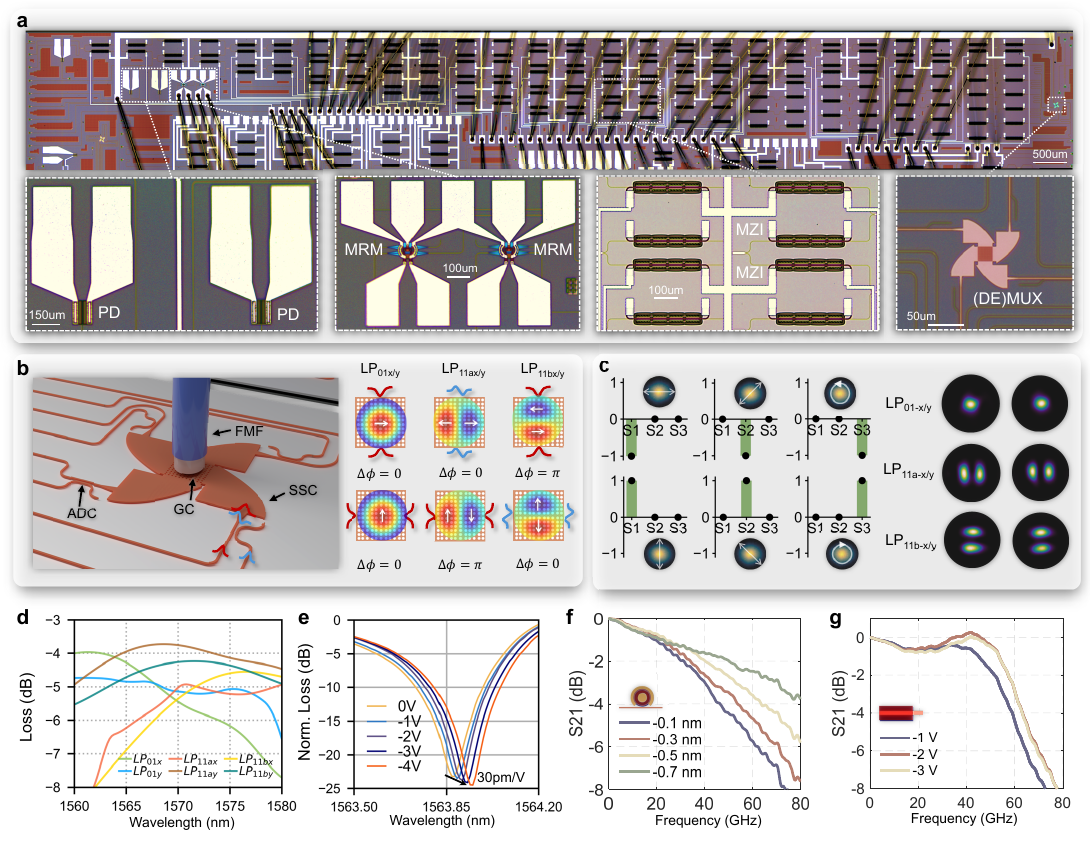}
  \caption{\textbf{Multidimensional silicon photonic I/O engine.} \textbf{a} Microscope images of the silicon photonic chip, showcasing zoomed-in views of high-speed germanium-on-silicon photodiodes (PDs), microring modulators (MRMs), Mach-Zehnder interferometers (MZIs) based optical signal processors, and MUX/DEMUX based on grating couplers (GCs). \red{Two input ports of the optical signal processor incorporate MRMs, while the output port employs a 50:50 on-chip splitter to tap light to the PD.} \textbf{b} Schematic and launching conditions for six orthogonal spatial and polarization beams. The few-mode fiber (FMF) is directly attached to the integrated MUX/DEMUX, which comprises a GC, a spot-size converter (SSC), and an adiabatic directional coupler (ADC). Transverse electric (TE) modes, specifically $\text{TE}_{0}$ and $\text{TE}_{1}$, are multiplexed into a common bus waveguide through the ADC. The SSC adjusts the mode field diameter before injection into the GC. Out-of-plane coupling is achieved with a precise vertical angle relative to the chip surface, directing the complex optical field into the FMF. Various spatial structures and polarizations can be excited, including linearly polarized (LP) modes $\text{LP}_{01x/y}$, $\text{LP}_{11ax/y}$, and $\text{LP}_{11bx/y}$. \textbf{c} Measured Stokes parameters and intensity profiles of optical beams diffracted by the integrated MUX, captured using a polarization analyzer works for quasi-Gaussian mode and an infrared camera. The launched beam structure is manipulated by tuning the integrated MZI-mesh with feedback control \textbf{d} Measured coupling loss spectra for the integrated MUX with a two-mode optical fiber to collect optical beam. \textbf{e} Measured optical transmission spectra under varying reverse bias voltages applied to the silicon MRM. \textbf{f} Measured electro-optical response of the MRM at different detuning wavelengths with a reverse bias voltage of -2 V. \textbf{g} Measured opto-electrical response of the germanium PD under various reverse bias voltages.}
  \label{fig:2}
\end{figure*}

Our photonic engine requires no pre‑calibration to launch or receive various spatial and polarization fiber modes. Instead,a feedback control loop adaptively addresses the dynamic and unpredictable transmission matrix between engines. Once the fiber link stabilizes, self‑configuration proceeds by iteratively adjusting the phase shifters within the optical mesh until a low‑crosstalk channel‑to‑channel transmission matrix is achieved. This process does not enforce predetermined spatial or polarization distributions. For example, a pair of quasi-Gaussian fiber modes can adopt different polarization distributions that remain orthogonal to each other, such as horizontal and vertical linear polarizations (HLP $\&$ VLP) or left- and right-circular polarizations (LCP $\&$ RCP). Consequently, the orthogonal wavefront set identified through self‑configuration extends beyond the standard linear polarization mode basis, forming complex speckle patterns that follow a random matrix rotation, as illustrated in Fig. \ref{fig:1}e. These simulated wavefronts were generated by randomly rotating the linearly polarized orthogonal basis vectors. In the multi-dimensional space created by the grating-fiber-grating system, self‑configuration searches for the optimal orthogonal physical channels, allowing light to follow the best natural paths rather than being constrained to a fixed set of optical routes. The resulting complex, orthogonal wavefronts arising from combined polarization and spatial diversity are distinct from the pure mode sets determined in free‑space configurations \cite{seyedinnavadeh2024determining}.

\section{Results} 
\subsection{Circuit Design and Characterization}
Fig. \ref{fig:2}a presents microscope images of the silicon photonic engine chip with microring modulators (MRMs), photodiodes (PDs), and optical signal processors. To save cost and reduce experiment complexity, only two MRMs and PDs are integrated on each photonic chip for IM/DD. External modulators and detectors are utilized for the other four physical channels to demonstrate its transparency against modulation formats and symbol rates. The complete set of orthogonal basis LP modes in a two-mode circular-core FMF can be excited by out-of-plane diffraction of transverse electric (TE) modes $\text{TE}_{0}$ and $\text{TE}_{1}$. Fig. \ref{fig:2}b shows the GC-based method as the coupling I/O and (DE)MUXs. The FMF is vertically aligned with the grating region. Eight single-mode nanowire channel waveguides are connected to adiabatic directional couplers (ADCs) and spot-size converters (SSCs) for the purpose of multiplexing the $\text{TE}_{0}$-$\text{TE}_{1}$ mode pair and facilitating changes in mode field diameter. Different mode excitations are achieved through the superposition of a pair of counter-propagating TE modes with varying relative phase differences. By employing the six supported LP modes as a basis set, any spatial or polarization state within a FMF can be represented as a linear combination of these modes. To demonstrate that, Fig. \ref{fig:2}c presents the measured Stokes parameters and intensity profiles of the launched LP\textsubscript{01} mode in different polarizations by adaptively tuning the MZI mesh. Spatial structure can also be manipulated by injecting different TE modes into the GC. The complete basis set of six LP modes is measured with an infrared camera, as shown in Fig. \ref{fig:2}c. \red{Details of the design and characterization are described in Supplementary Note S1.} Fig. \ref{fig:2}d shows the coupling loss spectra when various beams were collected by a FMF. \red{At approximately 1570 nm, a peak coupling efficiency of -3.7dB is achieved for the LP$_{11ay}$ mode, with a MDL of 1.8 dB across all supported spatial and polarization modes. Detailed theoretical predictions for the coupling losses of different LP modes are provided in Supplementary Note S1. The slight discrepancies between the simulated and experimental results primarily stem from inherent fabrication imperfections and fiber–chip misalignment.} In addition, the integrated MRM and PD were characterized for high speed data communication. The measured optical transmission spectra under varying reverse bias voltages for the MRM is illustrated in Fig. \ref{fig:2}e, showing a modulation efficiency of 30 pm/V. Fig. \ref{fig:2}f plots the electro-optical response of the MRM at various detuning wavelengths under a -2 V reverse bias. The 3-dB bandwidth of the MRM is around 52.8 GHz at a detuning wavelength of -0.5 nm. Fig. \ref{fig:2}g plots the opto-electrical response of the germanium-on-silicon PDs under different reverse bias voltages, showing a 3-dB bandwidth of around 60.4 GHz at a -2 V reverse bias voltage.

\begin{figure*}
  \includegraphics[width=0.95\linewidth]{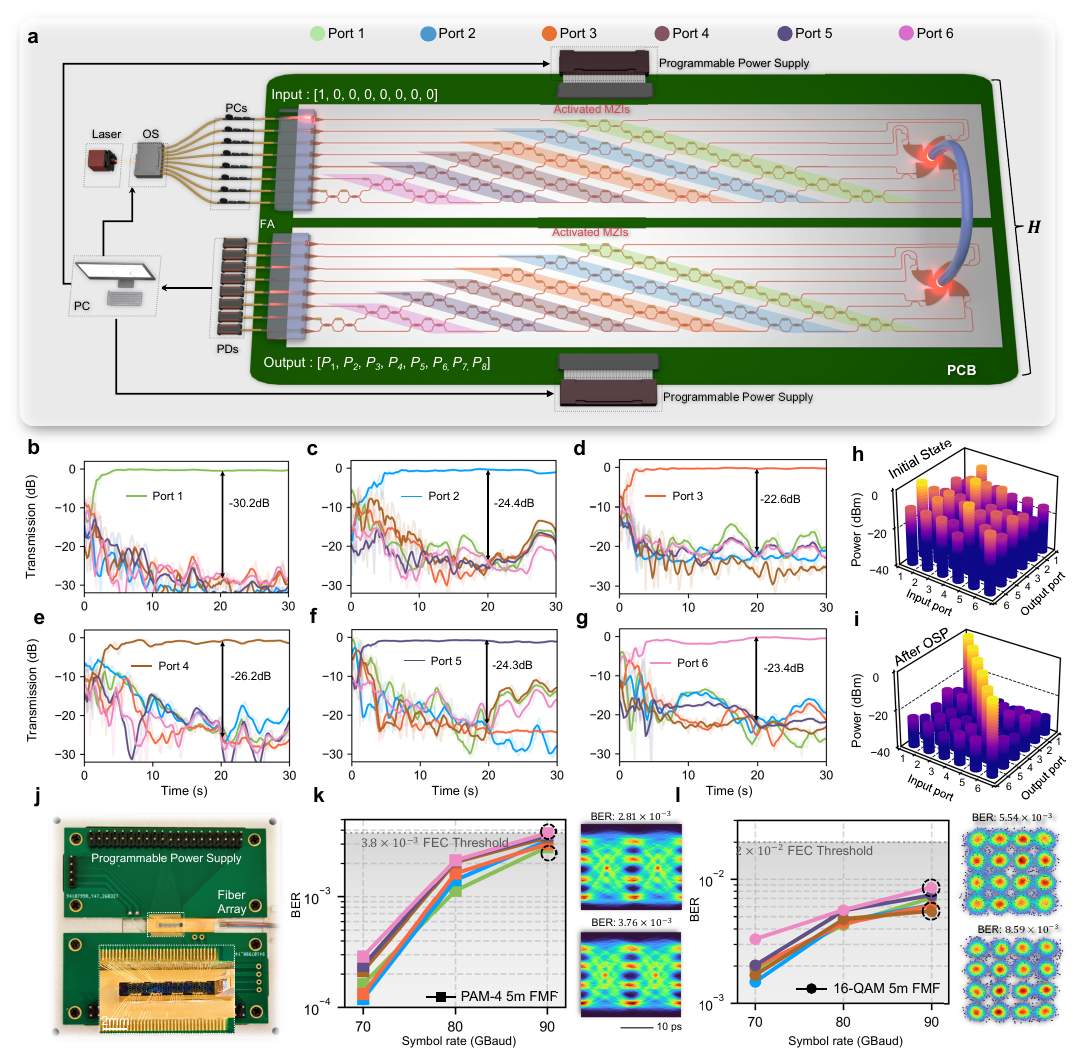}
  \caption{\textbf{Self configuration of MIMO-OSP processor for inter-chip channel-to-channel mapping.} \textbf{a} Experimental setup featuring two photonic chips mounted on a printed circuit board (PCB) and interconnected by a few-mode fiber (FMF). The input beam is selectively directed to one of the input ports using an optical switch (OS) controlled by a personal computer. Polarization controllers (PCs) are positioned before the fiber array (FA) due to the polarization-sensitive input grating coupler array. To address random channel transmission in the FMF, the MZI-based optical signal processors on both photonic chips are powered by a programmable supply. At the receiver photonic chip, outputs are directed to external PDs to provide feedback. The measured output optical power reflects one row of the transmission matrix \textbf{\textit{H}} for the entire optical system. \textbf{b-g} Normalized transmission evolution of six output ports during the $1\times6$ MIMO adaptive processor configuration. All six distinct input signals can be selectively routed to their corresponding output ports, achieving crosstalk levels ranging from -22.6 dB to -30.2 dB. The complete self-configuration process is accomplished in approximately 20 seconds. \textbf{h-i} Bar charts illustrating port-to-port transmission before and after activation of the optical signal processor. The observed disordered transmission distribution results from random spatial and polarization mixing. \textbf{j} Wired-bonded optical I/O chip with a fiber array and FMF attached. \textbf{k} Bit error rates (BERs) and corresponding eye diagrams of reference four-level pulse amplitude-modulation (PAM-4) signals when six ports are individually activated with a fiber transmission length of 5 meters. The optical signal symbol rate is varied from \red{70 to 90 Gbaud}. \textbf{l} BERs and corresponding constellation diagrams of reference 16 quadrature amplitude modulation (QAM) signals when six ports are individually activated with a fiber transmission length of 5 meters.}
  \label{fig:3}
\end{figure*}

\subsection{Photonic Engine Self-Configuration}

The self-configuration setup is illustrated in Fig. \ref{fig:3}a. Two silicon photonic engines were directly interconnected by a 5-m or 300-m FMF.  Light from an external laser was routed through an optical switch (OS) and a polarization controller (PC) before entering a fibre array (FA) attached to a grating‑coupler array on the transmitter engine. After passing through the chip-fiber-chip system, the optical powers at output ports $P_1 \sim P_8$ were measured by eight off-chip power meters. The phase shifters on two photonic engines were wire-bonded to printed circuit boards (PCBs) and driven by a programmable power supply under computer control. Although the optical signal processors were not calibrated, the power transmission matrix of the whole system can be estimated by sequentially injecting optical signals from the input ports. As illustrated in Supplementary Fig. S4b, each time an one-hot vector was sent to the system, and one row of the power transmission matrix $\mathbf{H}$ can be estimated. With the estimated transmission matrix $\mathbf{H}$, a particle swarm optimization (PSO) algorithm \cite{kennedy1995particle} was used to optimize the target signal processing function. Detailed MIMO-OSP self-configuration process is illustrated in the Supplementary Note S2.

\red{The $1\times6$ MIMO-OSP was first implemented to establish single-channel connectivity between the engines, ensuring successful point-to-point connectivity.} \red{In the experiment, the MZIs within the optical signal processor were selectively configured based on the power mapping between the input and output ports. Fig. \ref{fig:3}(a) illustrates that the specific MZIs were activated to establish the desired connections.} Figs. \ref{fig:3}b-\ref{fig:3}g illustrate the transmission iteration curves during self-configuration, establishing channel pairs between input port $i$ to output port $i$ ($1\leq i\leq 6$). Configuration from a random initial state took approximately 20 s on average. \red{The oscillations near the local optimum are attributable to the inherent stochasticity of the PSO algorithm, which can be effectively mitigated through an early stopping strategy.} The experimental results reveal crosstalk levels of -30.2 dB, -24.4 dB, -22.6 dB, -26.2 dB, -24.3 dB, and -23.4 dB for the six port pairs. Figs. \ref{fig:3}h and \ref{fig:3}i compare the power transmission matrix before and after the self configuration of the photonic processor. Fig. \ref{fig:3}j presents the photograph of the wired bonded optical engine mounted on a PCB with a fiber array attached. Through adaptive configuration for each port, six parallel waveguide ports can be aligned for inter-chip communication. Since signal processing occurs in the optical domain prior to photodetection, MIMO-OSP is agnostic to modulation formats and symbol rates. After adaptive configuration, external transmitters and receivers were used to generate and detect four-level pulse amplitude modulation (PAM-4) and 16-quadrature amplitude modulation (16-QAM) signals. Experimental setup for the data communication, along with a breakdown of optical power loss, is provided in Supplementary Note S3. \red{The symbol rate varied from 70 to 90 Gbauds, corresponding to data rates of 140 to 180 Gbps for PAM-4 and 280 to 360 Gbps for 16-QAM.} Figs. \ref{fig:3}k-\ref{fig:3}l present the measured reference bit error rates (BERs), eye diagrams, and constellation diagrams for each channel pair, confirming the feasibility of multidimensional, modulation‑format‑transparent inter‑chip communication.

\subsection{$\mathbf{N \times N}$ MIMO-OSP and Date Communication Experiments}
\begin{figure*}
  \includegraphics[width=0.95\linewidth]{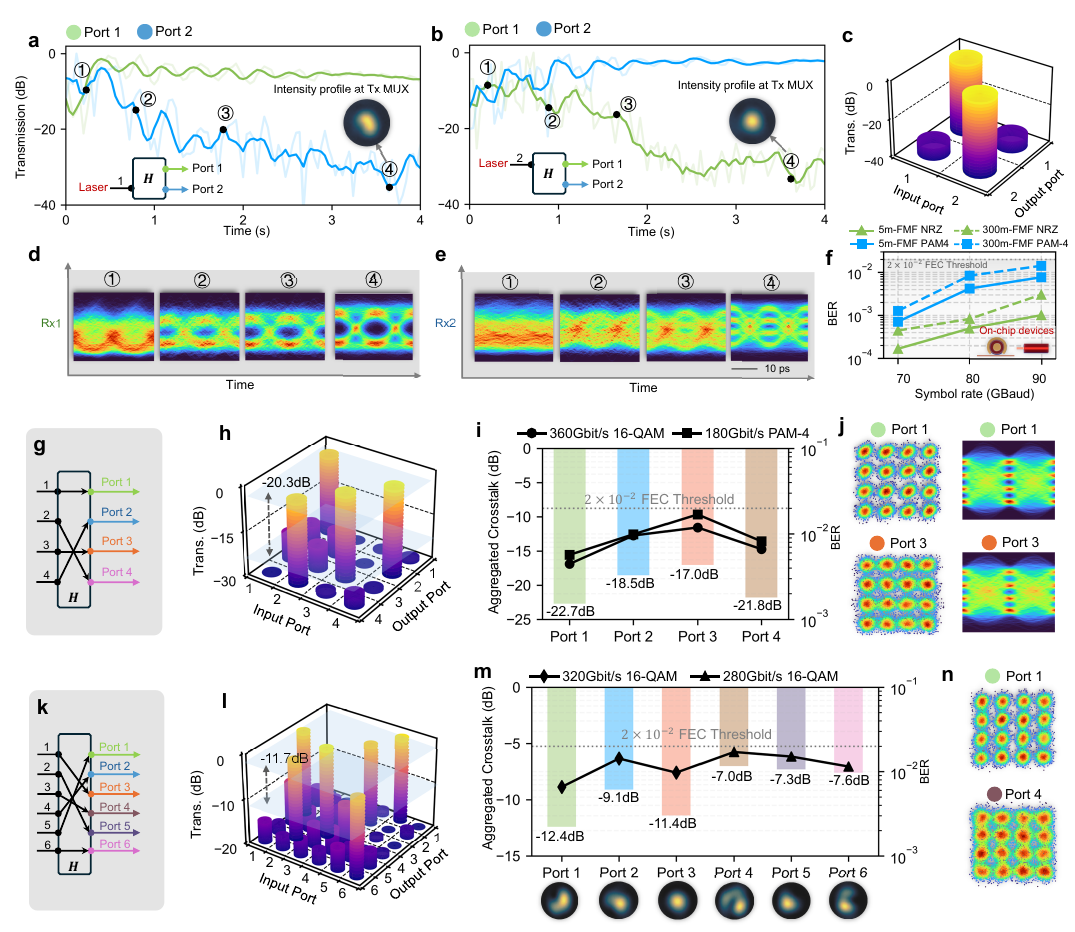}
  \caption{\textbf{Experimental results of MIMO-OSP for inter-chip optical communications and performance comparison with MIMO-DSP.} \textbf{a-b} Normalized transmission evolution of two output ports during the $2\times2$ MIMO adaptive processor configuration. The self configuration takes less than 4 second to automatically identify two concurrent physical channels for inter-chip communication. The insets show the intensity distribution of the corresponding beams emitted from the transmitter photonic chip. Each port is annotated with four distinct crosstalk levels: for port 1, the levels are 3.0 dB, -7.2 dB, -13.2 dB, and -37.1 dB; for port 2, the levels are 0 dB, -11.4 dB, -15.3 dB, and -33.1 dB. \red{\textbf{c} Transmission matrix after $2\times2$ MIMO self-configuration.} \textbf{d-e} Eye diagram evolution of the two concurrent channels during self configuration process. The mixed eye diagrams due to channel crosstalk are effectively mitigated by the implementation of the $2\times2$ MIMO-OSP. \red{\textbf{g} $4\times4$ MIMO self-configuration. \textbf{h} Transmission matrix after $4\times4$ MIMO self-configuration. \textbf{i} Aggregated crosstalk level and BER performance at each output port in $4\times4$ MIMO transmission. \textbf{j} Received constellations and eye-diagrams at the ports with highest and lowest crosstalk level. \textbf{k} $6\times6$ MIMO self-configuration. \textbf{l} Transmission matrix after $6\times6$ MIMO self-configuration. \textbf{m} Aggregated crosstalk level and BER performance at each output port in $6\times6$ MIMO transmission. The inset visualizes the speckle field generated by MUX diffraction. \textbf{n} Received constellations at the ports with highest and lowest crosstalk level. }   }
  \label{fig:4}
\end{figure*}

\red{In this section, the photonic engines were configured to support $N$ concurrent data channels via MIMO-OSP. We employ the similar strategy reported in \cite{miller2013establishing, seyedinnavadeh2024determining}, which involves the partial activation of MZIs within the mesh to reduce the optimization space (Supplementary Note S2). In the simplest case, a dual-channel concurrent transmission was enabled by a $2\times2$ MIMO-OSP scheme.} Figs. \ref{fig:4}a-\ref{fig:4}b show the transmission iteration curves observed during the self-configuration process. From a random initial state, inter-channel crosstalk was suppressed below -33.1 dB within just 4 seconds. The emitted beam intensity profiles were measured sequentially and are presented in the insets. The automatically identified orthogonal beam pairs exhibit no fixed spatial structures or polarization distributions as predicted in Fig. \ref{fig:1}e. \red{The resulting transmission matrix is shown in Fig. \ref{fig:4}c.} We then activated integrated silicon MRMs for IM/DD communication. Thermal heaters inside the MRMs were employed to finely tune their operational wavelengths to match the input light. \red{We applied 70-, 80-, and 90-Gbaud NRZ-OOK and PAM-4 modulation formats, detected using germanium PDs through ground-signal-ground-signal-ground (GSGSG) RF probes. BERs for both modulation formats after 5m and 300m of fiber transmission were presented by Fig. \ref{fig:4}f.} Figs. \ref{fig:4}d-\ref{fig:4}e present eye diagrams for 90‑Gbaud NRZ‑OOK and PAM‑4 at varying crosstalk levels during self‑configuration, where mixed eye patterns were transformed into clear open eyes after passing through the optimally configured optical network-on-chip, demonstrating its adaptive processing capability. Additionally, $2\times2$ MIMO-OSP was conducted for the remaining four ports, with results displayed in Supplementary Figure S6. Supplementary Video 1 shows real-time measurements of channel crosstalk and constellation diagrams throughout self-configuration. Furthermore, by leveraging the optical reciprocity, we demonstrated full-duplex inter-chip optical communication, as detailed in Supplementary Figure S7.

\begin{figure*}
  \includegraphics[width=0.95\linewidth]{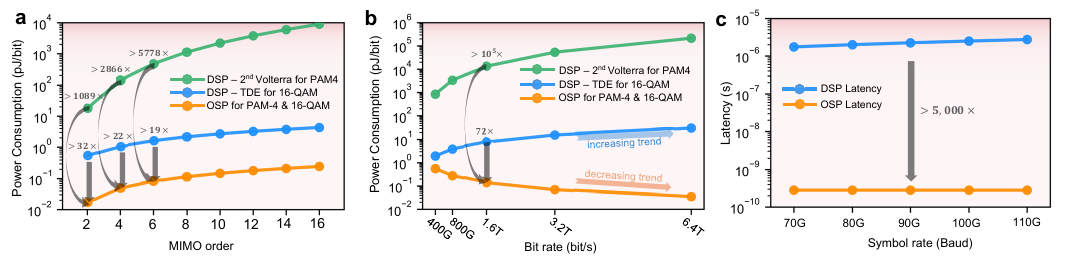}
  \caption{\red{\textbf{Comparison between MIMO-OSP and MIMO-DSP in the transmission over a 300-m FMF.} \textbf{a} Comparison of power consumption per bit for MIMO operation against processing order using the second-order Volterra nonlinear equalizer for 280-Gbps PAM-4, time-domain equalizer (TDE) for 280-Gbps 16-QAM, and OSP for both PAM-4 and 16-QAM. \textbf{b} Comparison of power consumption per bit for $6\times6$ MIMO operation against data throughput using Volterra nonlinear equalizer for PAM-4, TDE for 16-QAM, and OSP.  \textbf{c} Processing latency in $6\times6$ MIMO using DSP and OSP.}}
  \label{fig:5}
\end{figure*}

\red{The feasibility of 4 concurrent data transmission was further validated in our experiment utilizing off-chip transceivers. Fig.~\ref{fig:4}g shows a $4\times4$ MIMO transmission. The corresponding transmission matrix in Fig.~\ref{fig:4}h reveals that the measured non-cumulative crosstalk level for the four output ports are: -24.1 dB, -20.5dB, -20.3dB and -22.1dB, respectively. The aggregated crosstalk for each port was shown by Fig.~\ref{fig:4}i by summing optical power from all the other concurrent channels, ranging from -17.0 dB to -22.7 dB. High-speed data transmission is subsequently implemented. An aggregate rate of 1.44~Tbit/s or 0.72~Tbit/s can be reached using 360~Gbit/s 16-QAM signals or 180~Gbit/s PAM4 signals. Fig.~\ref{fig:4}i also illustrates the corresponding BER at each output port. The optimal and worst-case received eye and constellation diagrams are presented in Fig.~\ref{fig:4}j.}

\red{To push the performance limits, a $6\times6$ MIMO configuration is implemented, as shown in Fig.~\ref{fig:4}k. The transmission matrix (Fig.~\ref{fig:4}l) shows that the maximum non-cumulative crosstalk is below $-11.7$~dB despite the exponential growth of the search space with higher MIMO order. The measured  maximum non-cumulative crosstalk for the six output ports are: -18.3dB, -14.1dB, -14.8dB, -11.7dB, -12.8dB and -12.9dB, respectively. The corresponding aggregated crosstalk distribution is shown in Fig.~\ref{fig:4}m, ranging from -7.6 dB to -12.4 dB. For this $6\times6$ setup, Ports 1--3 can carry 320~Gbit/s 16-QAM signals, while Ports 4--6 can handle 280~Gbit/s 16-QAM signals, yielding a total throughput of 1.8~Tbit/s. The crosstalk-BER relationship, along with the speckle patterns generated from MUX grating, are detailed in Fig.~\ref{fig:4}m. Fig.~\ref{fig:4}n shows best‑ and worst‑case received constellations. Details of the simulation model for $N\times N$ MIMO self-configuration, along with the corresponding experimental iteration curves, are provided in Supplementary Note S5.}

We systematically quantify the overhead of electronic versus photonic MIMO processing during transmission over a 300-m FMF (Figs. \ref{fig:5}a–c). The detailed method can be found in Supplementary Note S4. Our comparison includes three MIMO equalizers: a second-order Volterra nonlinear equalizer for PAM-4, a linear time-domain equalizer (TDE) for 16-QAM, and an optical equalizer applicable to both 16-QAM and PAM-4, as reported in this study. For DSP electronics, we assumed the use of the latest 3-nm CMOS technology node \cite{cusmai20240}. The power consumption for MIMO-OSP is derived under worst-case conditions (all phase shifters were operated with a full $2\pi$ phase shift; see Supplementary Note S4). Fig. \ref{fig:5}a compares energy per bit versus MIMO order. \red{At a MIMO processing order of six, the power consumption overhead of the OSP is less than 0.1 pJ/bit, whereas the Volterra equalizer for IM/DD and TDE for coherent communications require 476.7 pJ/bit and 1.6 pJ/bit, respectively.} \red{As MIMO order increases, the Volterra equalizer remains >5,000 times more energy‑intensive than OSP. In contrast, the TDE and OSP show similar growth trends, but OSP consistently achieves about one order of magnitude lower energy consumption.}

\red{For the $6 \times 6$ MIMO equalizers, Fig. \ref{fig:5}b illustrates the power per bit against the data throughput per wavelength. At a data rate capacity of 1.6 terabit/$\lambda$/s, the TDE for 16-QAM consumes 7.8 pJ/bit and the Volterra equalizer for PAM-4 consumes $1.3 \times10^{4}$ pJ/bit, whereas the OSP consumes only 0.14 pJ/bit. This shows approximately 72-fold and $1 \times10^{5}$-fold improvements in energy efficiency, respectively.} For a fixed MIMO order, DSP power consumption improves at higher symbol rates, whereas OSP benefits from a reduced energy share per symbol, making its advantages even more pronounced at high symbol rates. Fig. \ref{fig:5}c compares processing latency versus symbol rate. MIMO‑OSP latency is simply the optical propagation time through the reconfigurable MZI network. In the case of DSP, we employed an optimistic estimation that accounts solely for parallelization latency, excluding internal data processing time. At a symbol rate of 64 Gbaud, the calculated latencies for DSP and OSP are $1.6~\mu s$ and 0.28 ns, which corresponds to a $5714\times$ reduction is latency overhead. Moreover, while DSP latency increases with symbol rate, OSP latency remains largely insensitive. These results demonstrate that MIMO‑OSP offers substantial advantages in energy efficiency, processing latency, and scalability for short‑reach inter‑chip communications.

\section {Conclusion and Discussion} 

\red{In this work, we have demonstrated a multidimensional silicon photonic engine that achieves a communication capacity exceeding 1 terabit/s per wavelength. By monolithically integrating optical transceivers, spatial and polarization (DE)MUXs, and optical signal processors on a single chip, the photonic engine can self-configure without prior characterization to identify optimal settings for two, four, and six orthogonal spatial and polarization channels. In comparison to state-of-the-art DSP, MIMO-OSP can be implemented at the speed of light by simply passing through a SVD-based reconfigurable optical mesh (<0.28 ns) with a small power consumption per bit (<0.1pJ/bit). Our analysis shows that at a throughput of 1.6 Tbps per wavelength, the energy efficiency improves by more than five orders of magnitude for IM/DD schemes and by 72‑fold for coherent solutions, while latency is reduced by 5,000‑fold. Furthermore, by exploiting the increased number of physical channels, full-duplex communication over a 300-meter few-mode fiber was enabled. These results establish a paradigm shift for optical engines in next-generation AI infrastructure and AI-driven data centers.}

\red{Despite the notable performance improvements demonstrated by MIMO‑OSP, self‑configuration remains a critical unresolved challenge in this work. The required reconfiguration frequency in real applications is dictated by the dynamics of environmental perturbations, primarily thermal fluctuations and mechanical drift. To maintain uninterrupted communication in real fiber networks, the photonic processor must adapt rapidly and precisely to these changing channel conditions. In our current implementation, the operating speed of thermo‑optic phase shifters is limited to a few megahertz \cite{parra2024silicon}. This could be accelerated to gigahertz rates by exploiting high‑speed electro‑optic effects \cite{li2023integrated, yuan20245, lu2025whispering}. Integrating feedback control electronics directly into the driving circuitry of the photonic chip presents a more effective strategy for reducing the time needed to determine the optimal communication channel set \cite{sacchi2025integrated, di2025high, zanetto2023time}. Moreover, the adaptive tuning capabilities of optical signal processors in high‑temperature environments, such as CPO or data centers, require further investigation. Our $N \times N$ MIMO self configuration results show that the parameter space expands significantly with matrix size. Consequently, simple evolutionary algorithms such as PSO become constrained by long search times and the risk of convergence to local optima. Precise and rapid training of on‑chip optical neural networks will therefore be essential \cite{zhou2020situ, zhou2020self}. The ability to swiftly and accurately perform SVD of an unknown multidimensional system transmission matrix will determine the feasibility of MIMO-OSP for practical deployment in future communication networks.}

\red{Our comparison of OSP with DSP highlights substantial advantages in energy efficiency, latency, and protocol transparency. To fully realize these benefits in future optical engines, the chip–fiber coupling interface must be further optimized in terms of efficiency, bandwidth, and dimensionality. Out‑of‑plane grating couplers offer flexibility and channel density, along with superior alignment tolerance, and their efficiency can be improved by stacking dielectric layers \cite{sacher2014wide, notaros2016ultra, zhou2025adaptive} or employing bottom reflection mirrors \cite{luo2018low, taillaert2004compact}. While current grating‑based (DE)MUXs exhibit limited bandwidth, recent edge‑coupled multimode multiplexers have demonstrated bandwidths exceeding 100 nm \cite{yi2024efficient, zhang2023ultra}. Using chip–fiber I/O devices as the (DE)MUXs for fibers could substantially reduce overall system footprint. Exploring a multidimensional chip–fiber interface is therefore a meaningful direction, as reconfigurable planar optics enables arbitrary manipulation of light over multiple degrees of freedom for advanced photonic applications. Recent advances in integrated structured‑light generators based on nanoantenna arrays offer a viable pathway to support a greater number of fiber modes \cite{butow2024generating, sharma2025universal}. The current prototype exhibits a millimeter-scale footprint, this form factor can be drastically reduced through compact MZI layouts or 3D photonic integration \cite{cao2026programmable, yang2017ultra}. Although the present work demonstrates terabit‑per‑second transmission via single‑wavelength multidimensional multiplexing and MIMO‑OSP, extending this architecture to WDM presents specific challenges. The primary bottleneck is modal dispersion in the fiber \cite{annoni2017unscrambling, yi2024unmixing}. Over practical distances, the transmission matrix evolves rapidly with wavelength, complicating unified broadband compensation using a single interferometric mesh. Dedicating separate meshes per wavelength is feasible but incurs large chip area and increased configuration power and time. Future efforts will needed to incorporate WDM techniques into our multidimensional optical engine.}

\section{Methods}\label{sec4}
\noindent\textbf{Optical I/O chip fabrication:} Silicon photonic chips were fabricated using an active multi-project wafer run at the Advanced Micro Foundry (AMF) Silicon Photonics Platform. The photonic integrated circuits employ a silicon-on-insulator (SOI) substrate. The top crystalline silicon layer is 220 nm thick; the buried oxide layer is 3 $\mu m$ thick. High-speed modulators utilize boron- and phosphorus-doped P-N junctions exploiting the plasma dispersion effect in silicon. Germanium epitaxial growth is implemented for high-speed photodetectors. Reactive-ion etching defines diffraction grating regions with an etch depth of 70 nm. Oxide cladding passivate the planar silicon waveguide devices. The reconfigurable Mach-Zehnder interferometer meshes are realized through integrated thermo-optic phase shifters employing titanium nitride (TiN) heaters above silicon waveguides.

\vspace{1em}
\noindent\textbf{Experimental setup for inter-chip communications:} The integrated photonic processor was powered by a multichannel programmable power supply (Time-Transfer T-MS128-12CV), which was automatically controlled by a personal computer receiving feedback from the optical power meter (Santec MPM-210H and MPM-215). The driving currents for the photonic processor were self-configured using a particle-swarm optimization algorithm. 

The few-mode fibers used in our experiments were sourced from OFS (Optical Fiber Solutions) The two-mode graded-index fiber has a core diameter of 16 $\mu m$ and a numerical aperture of 0.14. The optical fiber used in the data transmission experiment has a length of 300 meters. We employed an external tunable continuous-wave laser source (Santec TSL-770). An infrared camera (ARTCAM-991SWIR) and a polarimeter (Thorlabs PAX1000IR2/M) were used to characterize the beam structures emitted from the photonic chip.

The frequency response of silicon MRMs and germanium PDs were measured with a Lightwave Component Analyzer (Keysight LCA N4372E). For IM/DD, we utilized on-chip silicon MRMs and germanium PDs, along with commercially available high-speed thin-film lithium niobate Mach-Zehnder modulators (NOEIC MZ135-LN-110-C-H). The modulators were driven by an arbitrary waveform generator (AWG, Keysight 8199A) connected to RF amplifiers (SHF T850C). Optical signals were amplified by erbium-doped fiber amplifiers (EDFA, Amonics AEDFA-PA-35-B-FA) before being directed to the photodiodes (Coherent XPDV4121R). High-speed signals were captured using a real-time oscilloscope (Keysight UXR0702AP). For coherent communication, we employed an optical multi-format transmitter (IDPhotonics, OMFT Class 60) alongside a 70 GHz optical modulation analyzer (Keysight N4391C).

\medskip
\noindent \textbf{Additional Information}
\noindent Supplementary information is available in the online version of the paper.

\medskip
\noindent \textbf{Acknowledgments}
\noindent Y.T. acknowledges the support from the National Natural Science Foundation of China (No. 62305277), Guangdong Science and Technology Department (No.2024A1515012438), and Nansha District Key Area S\&T Scheme (No. 2024ZD007). P.M. and Y.T. acknowledge the support from Guangdong Science and Technology Department (No.2025B1212150003).

\medskip
\noindent \textbf{Author contributions}
\noindent H.C., Z.C., and Y.T. conceived the idea; Z.C., H.C., W.Z. designed the photonic integrated circuits; H.C., Z.C., W.Z. perform experiment assisted with K.L, M.Z., and Y.Y. ; K.L., Y.Y., and Y.C. designed the printed circuits board for wire bonding; H.C., W.Z., and Y.T. analyzed the data and prepared simulation models. H.C. and Y.T. wrote the manuscript assisted with all the other co-authors. C.H. and P.M. provided suggestions for experiments and revisions. P.M. and Y.T. funded the project. Y.T. supervised the project.

\medskip
\noindent \textbf{Competing interests} The authors declare no competing interests.

\medskip
\noindent \textbf{Data Availability} The data that support the findings of this study are included in the article and its supplementary information. Other data are available from the corresponding author upon request.

\section*{Reference}
\bibliographystyle{naturemag}
\bibliography{0_main_arxiv}

\end{document}